\documentclass[aps,preprint,showpacs,superscriptaddress,groupedaddress]{revtex4}  
\usepackage{graphicx}  
\usepackage{dcolumn}  
\usepackage{bm}
\usepackage{amssymb}
\usepackage{amsmath}
\usepackage{color}
\usepackage[normalem]{ulem}

\usepackage{todonotes}
\usepackage{comment}
\usepackage{hyperref}

\graphicspath{{fig/}{../fig/}}

\begin{document}

\title{Hamiltonian Memory: An Erasable Classical Bit}

\author{Roi Holtzman}
\affiliation{Department of Physics of Complex Systems, Weizmann Institute of Science, Rehovot, 76100, Israel}
\author{Geva Arwas}
\affiliation{Department of Physics of Complex Systems, Weizmann Institute of Science, Rehovot, 76100, Israel}
\author{Oren Raz}
\affiliation{Department of Physics of Complex Systems, Weizmann Institute of Science, Rehovot, 76100, Israel}
\email{oren.raz@weizmann.ac.il}
\date{\today}

\begin{abstract}
Computations implemented on a physical system are fundamentally limited by the laws of physics. A prominent example for a physical law that bounds computations is the Landauer principle. According to this principle, erasing a bit of information requires a concentration of probability in phase space, which by Liouville's theorem is impossible in pure Hamiltonian dynamics. It therefore requires dissipative dynamics with heat dissipation of at least $k_BT\log 2$ per erasure of one bit. Using a concrete example, we show that when the dynamic is confined to a single energy shell it is possible to concentrate the probability on this shell using Hamiltonian dynamic, and therefore to implement an erasable bit with no thermodynamic cost.
\end{abstract}

\maketitle

\section{Introduction}

In 1961, R. Landauer established a remarkable relation between information theory and thermodynamics, by arguing that an irreversible computation cannot be made without any energetic cost \cite{landauer1961irreversibility}. Landauer's principle is famously known as the statement that an erasure of one bit of information -- the hallmark of irreversible computations -- must dissipate at least $k_B T \log 2$ of heat, where $k_B$ is the Boltzmann constant and $T$ is the temperature of its surrounding environment. This bound is rooted in the dissipative dynamic, enforced by the contraction of the physical system's phase-space volume during the bit erasure. Such an operation cannot be done in an isolated, Hamiltonian dynamic, therefore Landauer concluded that implementing an erasable bit requires a dissipative system. By the second law of thermodynamics, the phase space volume reduction associated with the memory erasure must produce some dissipation, leading to the celebrated Landauer principle.

In the last few decades, Landauer's principle was refined and generalized to various cases. For example, it was generalized for a probabilistic erasure process, i.e. one that  only succeeds with some  probability \cite{maroney2009generalizing, gammaitoni2011beating}. Additional generalizations of Landauer's principle include other types of thermodynamic resources such as an angular momentum bath \cite{vaccaro2011information}, a bound for entropically unbalanced bits \cite{Bechhoefer2017PhysRevLett}, unifying the cost of erasing and measuring the bit \cite{Sagawa2009PhysRevLett,PhysRevLett.2016Crutchfield}, taking into account the mutual information between the bit and the bath \cite{sagawa2014Generalization}, $N$ state bit \cite{NBaseLogic2019generalization}, finite time erasure \cite{Becchoerfer2020arXiv1,Becchoerfer2020arXiv2} and others \cite{wolpert2019Review}. All these generalizations, however, rely exclusively on dissipative dynamics: following Landauer's argument, no energy conserving classical bit was suggested.

The memory technology used today is still far from approaching Landauer's bound. Nevertheless, in recent years several experiments used various physical systems to implement a dissipative bit that can be erased with an energetic cost which is close to Landauer's bound. These include colloidal particle in a trap \cite{Nature2012experimental,Bechhoefer2014prl,Ciliberto2015Experiment,Bechhoefer2017PhysRevLett,Bechhoefer2017PNAS}, nanomagnetic devices \cite{hong2016NanoMagneticExp,martini2016experimentalNanomagnetic,NanoMagnet2018NatPhys}, superconducting flux bit \cite{FluxBit2020nonequilibrium}, nuclear spin \cite{peterson2016experimental} and single molecule devices \cite{yan2018singleMoleculePRL,cetiner2020dissipation}. Apart from its practical implications, Landauer's principle is commonly used in theoretical studies to resolve seeming violations of the second law \cite{bennett1982thermodynamics,jarzynski2011modeling,bergli2014accuracy,parrondo2015Nature, marathe2010cooling}.

In this manuscript we present an exactly solvable and experimentally realizable example of a classical Hamiltonian system that can serve as a memory bit, which is erasable at no energetic cost. In this system, the erasure process maps an isolated part of the energy shell onto itself, assuring that the initial and final energies of the system are equal. In addition, the erasure maps most of the points on this isolated part into a small area, enabling to erase a bit of information without measuring it, and at no thermodynamic cost. We further show that this mapping can be done arbitrarily fast and to an arbitrary accuracy. This implies that contrary to a dissipative bit, such a Hamiltonian bit has no tradeoffs between duration, accuracy and energetic cost \cite{sagawa2014Generalization}. Crucially, the energy of the system has to be precisely known for this device to work.

\section{Hamiltonian Bit -- General Discussion}

\begin{figure*}
  \centering
  \includegraphics[scale=0.7]{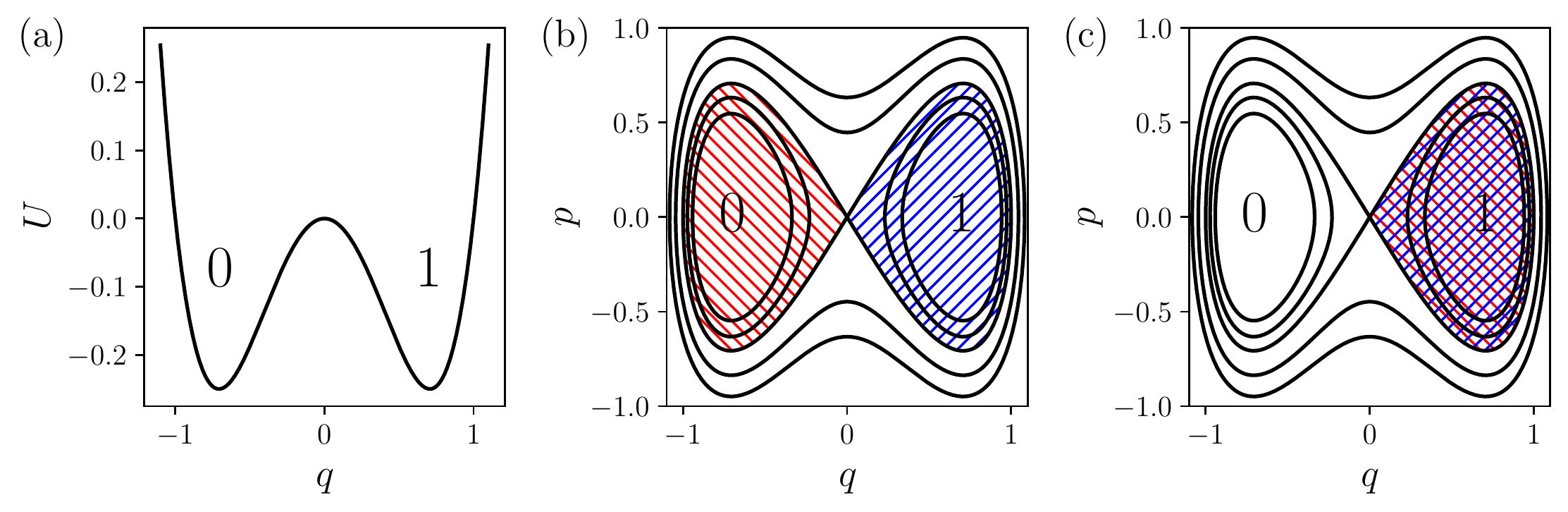}
  \caption{\texttt{restore-to-1} procedure performed on a Hamiltonian double well system. The right and left wells correspond to the `0' and `1' logical states, respectively. Initially, the bit is unknown -- this corresponds to a uniform phase space distribution in the two wells. \texttt{restore-to-1} brings the unknown bit to the `1' state.
  (a) The potential of a double well system.
  (b) Phase space of the initial configuration which corresponds to the state of an unknown bit -- the distribution is evenly split between the `0' (red) and `1' (blue) states.
  (c) The phase space distribution after a \texttt{restore-to-1} process should be concentrated at the state `1'. Both the initial red and blue regions in panel (b) must end in the red+blue region representing state '1', therefore the distribution must increase to compensate volume loss. This is incompatible with Hamiltonian dynamics, due to Liouville's theorem.}
  \label{fig:liouville}
\end{figure*}

To set the ground, let us first review the argument for the inapplicability of a pure Hamiltonian system as an erasable memory device. For a system to serve as a classical memory, there must be a mapping between its ``physical states'' and the ``memory states''. For simplicity we assume the memory to be a logical bit, whose memory states (often called logical states) are the `0' and `1' states. In addition, to serve as a useful bit the system must have an erasing mechanism. The logic operation of erasure takes any of the logical states, `0' and `1', and sets it to say, `1', at a high enough probability. Such a process is commonly referred as a \texttt{restore-to-1} procedure. Note that the procedure has to be performed without measuring the initial state of the bit, as by measuring, the information is copied to an external device which has to be erased too. Thus an effective erasure mechanism has to be able to erase a bit whose state is unknown. Physically, the \texttt{restore-to-1} procedure brings the system to a physical state associated with the logical state `1' (see Fig. \ref{fig:liouville}a-c). 

Let us consider such a \texttt{restore-to-1} procedure in Hamiltonian dynamics: As the initial logical state of the bit is unknown, the initial physical state of the bit is unknown as well. The unknown state can be represented by a uniform probability to be in any of the physical states associated with the two logical states `0' and `1' (see  Fig. \ref{fig:liouville}b). The erasure operation must concentrate this uniform probability distribution into the physical states associated with the logical state `1' (see Fig. \ref{fig:liouville}c). However, by Liouville's theorem, a Hamiltonian evolution cannot increase phase space probability \cite{goldstein2002classical}. Therefore, erasing a bit requires a non-conserving, dissipative dynamic, and is accompanied by a thermodynamic cost.

\begin{figure*}
  \centering
  \includegraphics[scale=0.7]{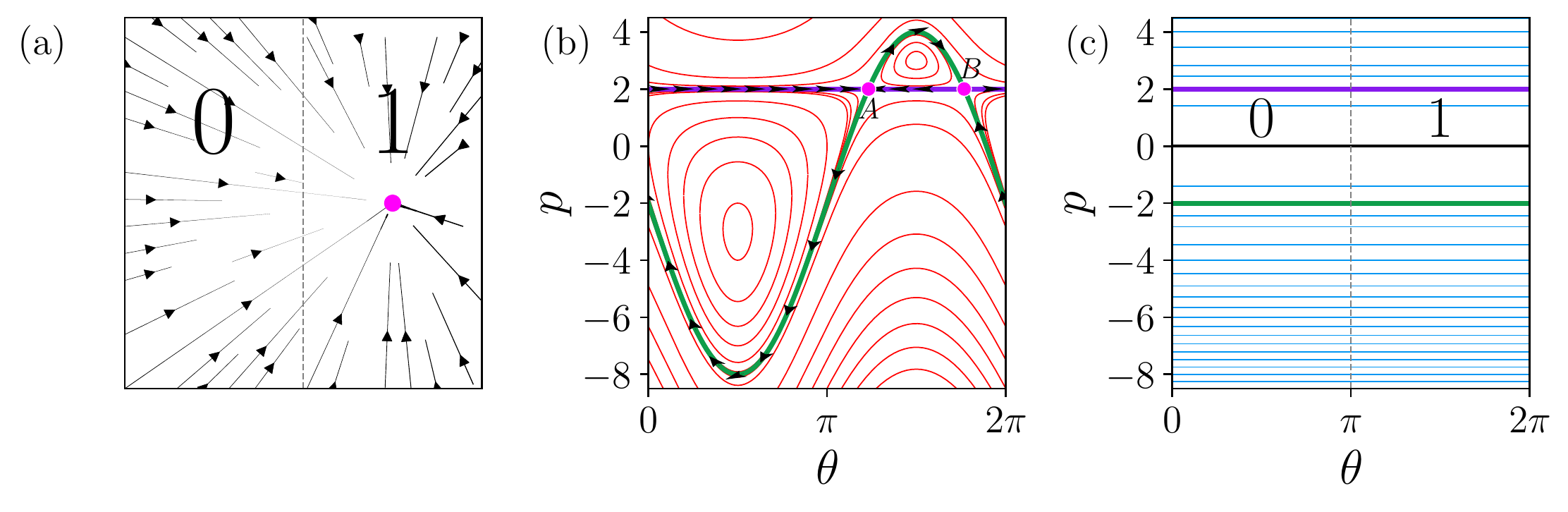}
  \caption{(a) The flow near an attracting fixed point (magenta) can serve as an eraser: initial conditions representing both the `0' state (left to the dashed line) and the `1' state (right) flow towards the state `1'.  (b) The phase space structure of $H_{ers}$ given in Eqs. (\ref{eq:9}), (\ref{eq:Hsq}) with $g=3$ and $E_0=2$. The fixed points $A$ and $B$ are on the same energy shell -- the green and violet contours. They have both attracting and repelling directions. On $\mathcal{E}_{ers}$ (the violet contour), all states flow towards the fixed point $A$, except state $B$. (c) The phase space structure of the Hamiltonian $H_0$, given in Eq. (\ref{eq:H_0}).  For $E>0$, energy shells are composed of two disconnected ergodic components, and can encode a bit of information -- the logical states `1' and `0' are marked. For $E=2$, the energy shell contains both $\mathcal{E}_0^+$ (violet) and $\mathcal{E}_0^-$ (green) ergodic components. $\mathcal{E}_0^+$ (the violet contour in (c)) is identical to $\mathcal{E}_{ers}$ (the violet contour in (b)).}
  \label{fig:Phase-Space-struct}
\end{figure*}

In what follows, it is shown that a pure Hamiltonian erasable memory device is nevertheless possible. We start with an abstract discussion, followed by a concrete example that demonstrates the idea. Consider a phase space flow that has a stable fixed point, attracting a large volume around it. The evolution under such a flow field can serve as an erasure mechanism, since after long enough time the state of the system is in the vicinity of the fixed point, regardless of where in the attracted volume it was initiated. If the vicinity of the fixed point represents the logical state `1', then the evolution under this flow generates a \texttt{restore-to-1} (see Fig. \ref{fig:Phase-Space-struct}a). Unfortunately, such a flow is incompatible with Hamiltonian dynamics: to satisfy Liouville's theorem and conserve phase space volume, a fixed point which is attractive in one direction must be repelling in some other direction (see Fig. \ref{fig:Phase-Space-struct}b). However, if the evolution of the system is confined to the specific manifold which is attracted to the fixed point, then the corresponding Hamiltonian can serve as an eraser, even though phase space volume is conserved. Such a confinement can be generated by a conserved quantity -- in what follows we use conservation of energy. We denote the manifold of states that are attracted to the fixed point, which is part of the energy shell, by $\mathcal{E}_{ers}$ (e.g. the violet contour in Fig. \ref{fig:Phase-Space-struct}b). For a time independent Hamiltonian, a system that is initiated on $\mathcal{E}_{ers}$ stays on it. Therefore, a Hamiltonian with such a fixed point, denoted by $H_{ers}$, can serve as an eraser -- provided that all of the physical states corresponding to the memory are confined to $\mathcal{E}_{ers}$.

So far we have described how the erasure is performed. Let us now turn to the physical description of the bit's steady state. When no operation is made on the bit, it evolves under a different, time independent Hamiltonian, $H_0$. The evolution under $H_0$ is confined to an energy shell of $H_0$ that corresponds to the initial condition.  This energy shell might include several disconnected surfaces, and the evolution of the system is confined to the specific surface based on the initial condition. These disconnected surfaces composing the energy shell are referred to as \emph{ergodic components}, and we denote them by $\mathcal{E}_0^i$ (see Fig. \ref{fig:Phase-Space-struct}c for an example). To perform erasure, the Hamiltonian is changed from $H_0$ to $H_{ers}$. We have seen that $H_{ers}$ erases states that are confined to $\mathcal{E}_{ers}$, thus the evolution under $H_0$ must also be confined to $\mathcal{E}_{ers}$. Therefore, $H_0$ must have one of its ergodic components coinciding with $\mathcal{E}_{ers}$ (compare the violet curves in Fig. \ref{fig:Phase-Space-struct}b, \ref{fig:Phase-Space-struct}c). For reasons that will become clear later, we denote this ergodic component by $\mathcal{E}_0^+$.   Lastly, the different states in $\mathcal{E}_0^+$ must encode the logical bit in a way that is not altered by $H_0$, and such that the vicinity of the fixed point of $H_{ers}$ encodes the logical state `1'.

Let us summarize the construction described above. The system is initially prepared on a specific ergodic component $\mathcal{E}_0^+$ of $H_0$. The state of the system in $\mathcal{E}_0^+$ encodes the logical bit, and the evolution under $H_0$, which is confined to $\mathcal{E}_0^+$, conserves this logical state. To erase the bit, $H_{ers}$ is  switched on for some time, until large enough portion of $\mathcal{E}_0^+$ is concentrated by the evolution under $H_{ers}$ near the fixed point, and the corresponding logical state is known to a high enough accuracy. At this point, $H_{ers}$ is switched off, and the system continues to evolve under $H_0$.

To generate such a bit, two different experimentally realizable Hamiltonians with the relevant overlap between their energy shells are required. Consider first the simplest form of a Hamiltonian,
\begin{equation}\label{Eq:SimpHamilt}
    H = \frac{p^2}{2} + U(q),
\end{equation} 
where we set the mass $m=1$  and $U(q)$ is a position dependent potential. Two different Hamiltonians $H_0$ and $H_{ers}$ cannot have the desired overlap between their energy shells when both of them are of the form in
Eq. (\ref{Eq:SimpHamilt}): the contribution of $p^2/2$ is identical in both Hamiltonians, therefore on $\mathcal{E}_0^+$ the two Hamiltonians must have the same $U(q)$ up to a constant, and both Hamiltonians represent exactly the same physics at this energy and this range of $q$. For the two Hamiltonians to be different and nevertheless have the desired overlap between their energy shells, we consider Hamiltonians that include magnetic fields,
\begin{equation}\label{Eq:ExperimentalHamiltonForm}
    H = \frac{\Big(p-A(q)\Big)^2}{2} + U(q).
\end{equation} 
Note that the Hamiltonian in Eq. (\ref{Eq:SimpHamilt}) is symmetric with respect to sign changes of $p$, whereas the one in Eq. (\ref{Eq:ExperimentalHamiltonForm}) is not. 

\section{An Illustrative Example -- a Particle on a Ring}
\label{sec:rotator-example}

Next, we consider a specific example that illustrates the idea presented above: a particle moving on a ring, with the coordinate $\theta \in [0,2\pi)$ denoting the angle of the particle on the ring. The Hamiltonian of the particle is given by
\begin{equation}
\label{eq:H_0}
H_0(\theta,p) = \frac{p^2}{2}.
\end{equation}
 The phase space of this system is very simple. Each energy shell (except $E=0$) is composed of two disjoint ergodic components corresponding to clockwise ($p>0$) and counter-clockwise ($p<0$) rotations (see Fig. \ref{fig:Phase-Space-struct}c). In what follows we consider the $p>0$ ergodic component of the energy shell, defined as $\mathcal{E}_0^+= \left\{ (\theta,p): H_0(p) = E_0;\; p>0 \right\}$. We denote $p_0 = \sqrt{2E_0}$, so a state with momentum $p_0$ belongs to the ergodic component $\mathcal{E}_0^+$.

To encode one bit of information in the physical state of the particle, which constantly evolves in time, we exploit the periodicity of the particle's motion. The period depends on the exact value of the energy of the system, $E_0$. Therefore, for $\mathcal{E}_0^+$ with a period  $\tau = 2 \pi / p_0$, we associate a logical state according to the physical state of the system at stroboscopic time intervals $\tau$, namely at an integer multiplication of $\tau$. A physical state $(\theta,p)$ in a stroboscopic time is assigned a logical state using, for example, the following mapping (see Fig. \ref{fig:Phase-Space-struct}c):
\begin{align}
\label{eq:bit-mapping}
  Logic(\theta, p) = \begin{cases}
    0 &\mbox{if } \theta \in [0, \pi) \\ 
    1 &\mbox{if } \theta \in [\pi, 2\pi).
  \end{cases} 
\end{align}

The Hamiltonian $H_0$ controls the evolution of the bit when no operation is performed on it, and it conserves the encoded logical state. To control the bit, and specifically to perform a \texttt{restore-to-1} procedure, a different Hamiltonian must be applied on the system for a finite time. \texttt{restore-to-1} means that most initial states $(\theta_i,p_i) \in \mathcal{E}_{0}^+$ are mapped to a final state $(\theta_f,p_f) \in \mathcal{E}_0^+$ that encodes the bit 1, i.e. $Logic(\theta_f,p_f) = 1$. Note that all final states $(\theta_f,p_f)$ must belong to the same energy shell $\mathcal{E}_0^+$, otherwise the period of the particle differs from the stroboscopic time used in the definition of the logical state of the bit, causing the bit to decohere and slowly lose its information.

To erase the bit, we need to find a Hamiltonian, $H_{ers}(\theta,p,t)$, that has three properties: (i) It maps the specific ergodic component   $\mathcal{E}_0^+$ into itself in a finite time; (ii) It concentrates the uniform distribution on this energy shell, and (iii) It is experimentally plausible. In what follows, we refer to a Hamiltonian as having property (iii) if it has the form given in Eq.(\ref{Eq:ExperimentalHamiltonForm}). Although there are many  Hamiltonians that satisfy all these constraints, finding one of them is not a trivial task.  Next, we show a systematic method to find such a Hamiltonian, by interpolating between $H_0$ and a different Hamiltonian that shares the ergodic component $\mathcal{E}_0^+$ with $H_0$.

Let us formulate the conditions (i) and (ii) in terms of the of the $\theta, p$ phase space coordinates of the system. Condition (i) is satisfied if any initial state $(\theta_i, p_i = \sqrt{2 E_0}) \in \mathcal{E}_{0}^+$ evolves under $H_{ers}$ to a point $(\theta_f, p_f)$ with the same momentum value, namely $p_f = \sqrt{2 E_0}$. A necessary condition for (ii) is that the coordinate velocity $\dot{\theta}$ is a function of $\theta$. Otherwise, all the coordinates evolve with the same rate, and the uniform distribution remains a uniform distribution.

To generate probability concentration we add a term to the Hamiltonian, denoted by $H_{con}$. The full erasure Hamiltonian is given by
\begin{equation}
\label{eq:9}
H_{ers}(\theta,p,t) = H_0(p) + g(t)H_{con}(\theta,p),
\end{equation}
where $g(t)$ is a function that ramps the concentration term  $H_{con}$ on and off in a continuous manner and controls the concentration magnitude.  With this Hamiltonian the equations of motion are
\begin{align}
  \dot{p} & = - \frac{\partial H_{ers}}{\partial \theta} = - g(t) \frac{\partial H_{con}}{\partial \theta} \label{eq:p-dot-Hers}\\
  \dot{\theta} & = \frac{\partial H_{ers}}{\partial p} = p + g(t) \frac{\partial H_{con}}{\partial p} .  \label{eq:q-dot-Hers}
\end{align}
One way to meet condition (i) is to have $\frac{\partial H_{con}}{\partial \theta} \Big|_{p_0} = 0$, which implies that $p_0$ is constant. In this case, $\mathcal{E}_0^+$ is part of the energy shell of $H_{ers}$ too.

A choice for $H_{con}$ that satisfies conditions (i), (ii) and (iii) is
\begin{equation}
\label{eq:Hsq}
H_{con}(\theta,p) = \left( p - \sqrt{2 E_0 } \right) \sin(\theta).
\end{equation}
On $\mathcal{E}_0^+$ we have $p=\sqrt{2 E_0}$, so $H_{con}=0$. Therefore it is $\theta$ independent on $\mathcal{E}_0^+$, and thus $p$ is conserved on this ergodic component (see Eq. (\ref{eq:p-dot-Hers})). In addition, $H_{con}$ has only a linear term in $p$ which can be brought to the the form of Eq. (\ref{Eq:ExperimentalHamiltonForm}), and therefore it fulfills condition (iii).

\begin{figure}
  \centering
  \includegraphics[scale=0.75]{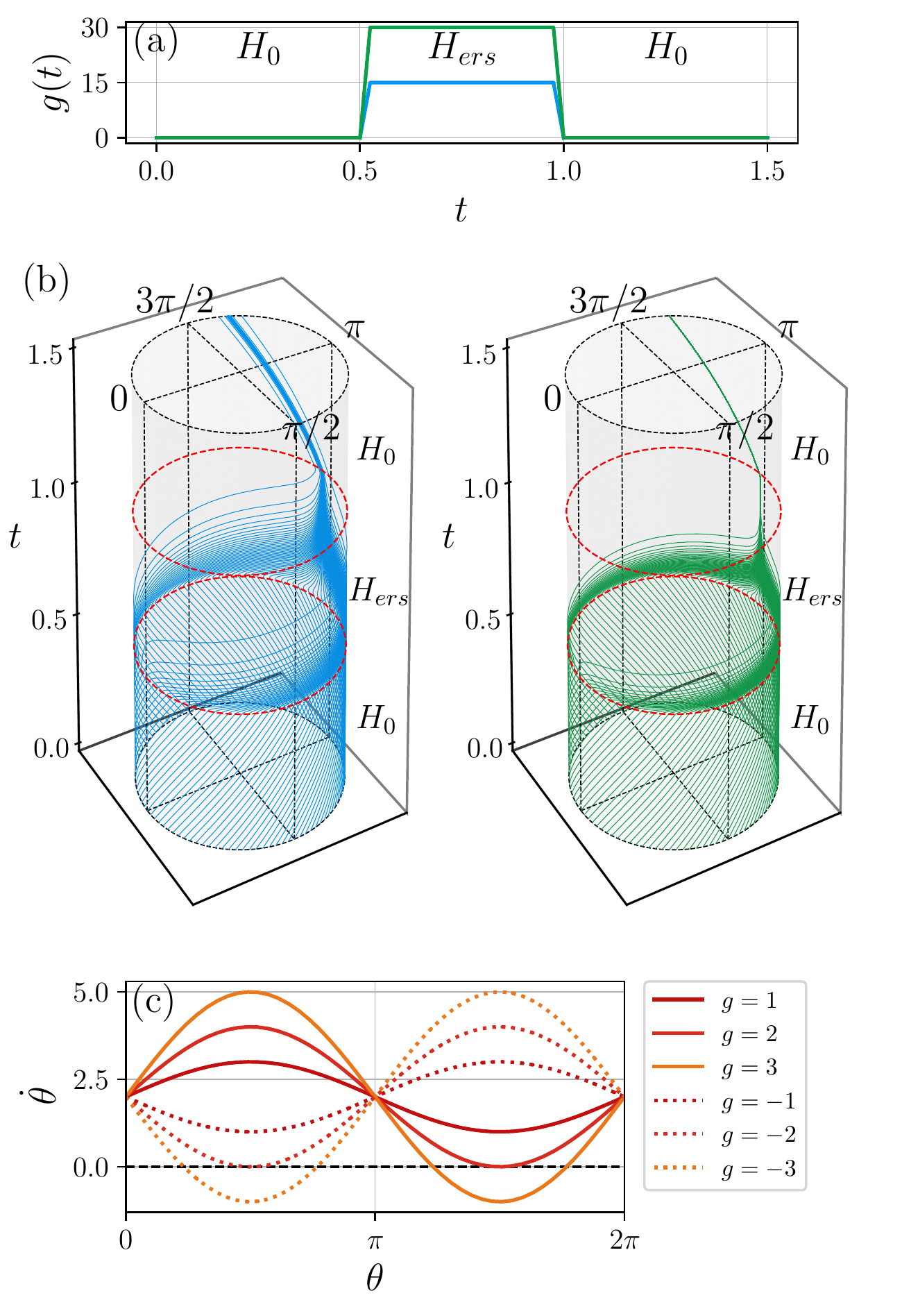}
  \caption{(a) The protocols $g(t)$ used in $H_{ers}(t)$, for the evolution in (b). (b) Evolution under $H_{ers}(t)$ of 100 uniformly distributed initial angles. The angle $\theta$ is represented as the angle on the unit circle, and the time evolution is along the vertical axis. For $0\leq t<0.5$ the system evolves under $H_0$, and therefore the distribution remains uniform. From $t=0.5$ to $t=1$, a \texttt{restore-to-1} protocol is applied, with $g(t)$ shown in (a). The left panel corresponds to maximal value $\max g(t)=15$ (blue), and the right panel to $\max g(t) = 30$ (green). At the end of these protocols, at $t=1$, all the angles are extremely concentrated around one value $\theta_s$. Then the bit continues to evolve under $H_0$ for $1 \leq t<1.5$, during which the concentrated distribution rotates at a constant velocity. Evidently, larger $g$ results in a faster and more accurate erasure. (c) $\dot{\theta}$ as a function of $\theta$ given in Eq. (\ref{eq:q-dot-on-shell}), for a few values of $g$. For $g > \sqrt{2E_0} = 2$, a stable and an unstable fixed points exists and the microcanonical distribution is singular.}
  \label{fig:cylinder-evolution}
\end{figure}

Let us next show that $H_{con}$ provides the desired probability concentration, as required by condition (ii). The equations of motion for the full erasure Hamiltonian on the specific ergodic component $\mathcal{E}_0^+$ for the specific choice of $H_{ers}$ are given by
\begin{align}
  \dot{p} |_{p_0} & = 0 \label{eq:p-dot-on-shell} \\ 
  \dot{\theta} |_{p_0} & = p_0 + g(t) \sin \theta. \label{eq:q-dot-on-shell}
\end{align}
The evolution of any angle $\theta_i \in \mathcal{E}_0^+$ is clear from Eq. (\ref{eq:q-dot-on-shell}). For $g(t) > p_0$, $\dot{\theta}$ vanishes twice at $\theta_s$ and $\theta_u$ which correspond to stable and unstable fixed points of the dynamic of the angles (see Fig. \ref{fig:cylinder-evolution}c for its plot at several values of $g$). Therefore, for $g(t) > p_0$ the dynamic has a fixed point on $\mathcal{E}_0^+$, and all angles flow towards the stable angle $\theta_s$. Moreover, for large enough $g(t)$, $\dot{\theta}$ scales with $g(t)$ for most values of $\theta$. Thus, most initial conditions flow very rapidly to $\theta_s$. This ensures that almost all initial angles converge to $\theta_s$ provided that the duration protocol $\mathcal{T}_{ers}$ and the magnitude of $g(t)$ are both large enough. Moreover, it shows that there is no fundamental bound on the rate of probability concentration, and that the concentration, namely the probability to find the system in a state that corresponds to the desired bit, can be arbitrarily high. Indeed, for any arbitrarily small erasure time $\mathcal{T}_{ers}$, one can choose large enough $g(t)$, such that almost all initial angles $\theta_i \in \mathcal{E}_0^+$ are mapped arbitrarily close to the stable angle $\theta_s \in \mathcal{E}_0^+$. However, not \emph{all} initial angles can be mapped to $\theta_s$, since the ring topology of the shell is invariant by the dynamic, and thus cannot be torn. Nevertheless, the fraction of initial angles mapped close to $\theta_s$ can be increased arbitrarily by increasing $g(t)$. Lastly, we note that large enough $g(t)>0$ results in $\theta_s>\pi$, whereas small enough $g(t)<0$ results in $\theta_s< \pi$, therefore by controlling the sign of $g(t)$ both \texttt{restore-to-1} and \texttt{restore-to-0} can be implemented.

To show the convergence of most initial conditions, 100 uniformly spaced initial angles $\left\{ \theta_j \right\}_{j=1}^{100} \in [0, 2\pi)$ were chosen in Fig. \ref{fig:cylinder-evolution}, and evolved according to a specific protocol: for $0\leq t<0.5$ the system evolved under $H_0$ with $E_0=2$, so the angles evolved uniformly. During $0.5\leq t<1$ the erasure is performed by $H_{ers}(t)$ as defined above. The specific choice of $g(t)$ is shown in the upper panel of the figure. Lastly, the bit evolved again under $H_{0}$ for $1 \le t<1.5$. The left and right central panels show the evolution of $\theta_j(t)$, with different amplitudes for $g(t)$. In agreement with the analysis above all of the trajectories end up with almost the same final (stable) angle $\theta_s$. Moreover, the rate of convergence increases with the magnitude of $g(t)$.

In terms of the angle distribution, $P(\theta)$, the erasure procedure takes the initial microcanonical uniform distribution, $P_{\mu c}(\theta) = 1/2\pi$, and maps it to a continuous distribution $P_f(\theta)$ which is concentrated around $\theta_s$. Clearly, the Shannon entropy associated with this distribution, given by
\begin{equation}
\mathcal{S}(P) = - \int\limits_0^{2\pi} P(\theta) \log P(\theta) d\theta,
\end{equation}
decreases during the protocol. In the specific protocols implemented in Fig. \ref{fig:cylinder-evolution}, the changes are $\Delta \mathcal{S} = -5.69$ for $\max g(t)=15$, and $\Delta \mathcal{S} = -12.83$ for $\max g(t)=30$. This decrease in the information entropy comes with no energetic cost, as the initial and final states of the system are on the same energy shell, so no work is done on the system. Moreover, the protocol can be repeatedly applied on as many bits as needed without any measurement or any other type of thermodynamic cost.

Does the above construction violate the second law? To see why this is not the case, we note that the Hamiltonian dynamic is reversible. In other words, even after the erasure process, it is in principle possible to reverse the dynamic and find the initial state of the system. Therefore, no physical entropy is generated in the erasure process. However, the logical state is irreversible: if only the logical state after the erasure is known, namely that the system represents the logical state `1', then there is no way to know what was the previous state of the system. Logical irreversible process with a reversible dynamic was already pointed out by Sagawa in \cite{sagawa2014Generalization}, where he argued that the combination of the bit and a thermal bath together is a closed system with a reversible dynamic.

\section{Finite Width Shell}
\label{sec:finite-widht-shell}

The results presented above were made possible by tailoring $H_{ers}$ to the specific ergodic component $\mathcal{E}_0^+$ (note the dependence of $H_{con}$ on $E_0$ in Eq. (\ref{eq:Hsq})), and only bits on it are erased at no energetic cost. Therefore, one has to know $E_0$ precisely. In classical mechanics, a single system has a unique value for its energy, and there is no limit on how accurately the energy can be known. Since the system is isolated and the \texttt{restore-to-1} protocol conserves the known energy, it can, in principle, be kept constant at all times. However, in reality there is always some finite precision limiting the accuracy at which the energy is known. Such finite accuracy corresponds to a thick energy shell, on which the uniform probability distribution cannot be increased, and compressing the distribution in the $\theta$ direction results in spreading of the distribution along the $p$ coordinate, and thus along the energy. Therefore, with repeated erasures the accuracy of the energy decreases, and the lifetime of the bit decreases accordingly.

\section{Conclusions}
\label{sec:conclusions}

To conclude, we have shown with a simple solvable example, that a well isolated energy conserving system can serve as a mutable memory. If the energy of the system is known precisely, then there is \emph{no} thermodynamic cost in erasing the bit. In our construction, we used a Hamiltonian dynamic that can be implemented in experiment, as it consists of a kinetic energy term, a term that is proportional to the momentum multiplied by a function of position -- such a term can in principle be implemented using magnetic fields, and a potential which is a function of the coordinate alone. Alternatively, the linear momentum term can be removed to yield a time dependent Hamiltonian with a standard kinetic term and a position-only dependent potential by a simple transformation as shown in \cite{jarzynski2017fast}. However, we ignored many other practical limitations, as the maximal force that can be applied on the system or the rate at which the potentials can be changed in space and time. We also ignored limitations imposed by quantum mechanics, e.g. the uncertainty principle which implies minimal width to any energy shell. 

The construction used in this manuscript keeps the state of the system on a specific energy shell at all time, and uses a fixed point in $H_{ers}$ to concentrate the probability distribution. We note that these are not essential to construct an erasure: concentration of probability occurs even if $H_{ers}$ does not have a fixed point as long as $\mathcal{E}_0^+$ coincides with $\mathcal{E}_{ers}$ and the microcanonical distribution on $\mathcal{E}_0^+$ is different for $H_0$ and $H_{ers}$. Alternatively, there is no need for the dynamic to stay confined on $\mathcal{E}_0^+$ at all times when $H_{ers}$ operates: it is enough that $H_{ers}$ maps $\mathcal{E}_0^+$ onto itself at a specific time.


\end{document}